# Quantum Control of Molecular Gas Hydrodynamics

S. Zahedpour, J. K. Wahlstrand and H. M. Milchberg*

*Institute for Research in Electronics and Applied Physics*
*University of Maryland, College Park*
*milch@umd.edu

We demonstrate that strong impulsive gas heating or heating suppression at standard temperature and pressure can occur from coherent rotational excitation or de-excitation of molecular gases using a sequence of non-ionizing laser pulses. For the case of excitation, subsequent collisional decoherence of the ensemble leads to gas heating significantly exceeding that from plasma absorption under the same laser focusing conditions. In both cases, the macroscopic hydrodynamics of the gas can be finely controlled with ~40 fs temporal sensitivity.

Significant hydrodynamic perturbation of solids, liquids, and non-dilute gases by nonlinear absorption of intense laser pulses typically proceeds by localized plasma generation, which provides the pressure and temperature gradients to drive both mass motion and thermal transport. This is typically assumed to be the case for femtosecond filaments in gases, where depletion of the laser pulse energy due to absorption limits their ultimately achievable length [1] and where the thermal energy deposited in the gas can result in sound wave generation [2-9] followed by a gas density depression or 'hole' that can persist on millisecond timescales [5-7]. Recently it was shown that this density hole can affect filamentation at kilohertz repetition rates by acting as a negative lens [5] and can steer filaments [6]. It may even play an important role in filament-triggered electrical discharges [3, 4]. New applications such as high average power laser beam guiding and remote generation of lensing structures in the atmosphere [7] can be enabled by control of energy deposition in gases by femtosecond laser pulses.

Previous work on energy deposition by filamentation has emphasized laser absorption due to atomic or molecular ionization and heating of free electrons [3, 4, 5, 10]. In this paper, we first show that molecular rotational heating is the dominant source of energy absorption in air filaments produced by single pulses. We then show that significantly greater gas heating can be generated by coherently and resonantly exciting a molecular rotational wavepacket ensemble by a sequence of short non-ionizing laser pulses separated by the rotational revival period [11, 12]. By 'wavepacket', we mean the coherent superposition of rotational eigenstates $|j,m\rangle$ excited by the laser pulse(s), where $j$ and $m$ are quantum numbers for rotational angular momentum and for the component of angular momentum along the laser polarization. Gas heating occurs by collisional de-excitation and decoherence of the ensemble, leading to significant hydrodynamic response. Gas heating can be equivalent to that driven by a filament plasma heated up to ~100 eV, greatly in excess of typical filament plasma electron temperatures of ~ 5 eV. Moreover, we show that it is possible to deplete a population of rotationally excited molecules before the wavepacket collisionally decoheres, suppressing gas heating. These results point to new ways of precisely controlling gas density profiles in atmospheric propagation [7], and have practical implications for schemes using pulse trains to enhance supercontinuum generation, filament length and plasma density [13, 14], and THz amplification [11]. Other novel extensions are also suggested. Isotope-selective pumping of rotational population was measured recently at low temperatures and pressures [15]. The interferometric technique employed here, combined with optical centrifuge techniques [16, 17] or chiral pulse trains [18], could find use in studying laser-induced gas vortices [19].

Here, laser excitation/de-excitation of the molecular ensemble is monitored by direct interferometric measurement of the gas density depression produced by subsequent heating of the gas. A short laser pulse is absorbed by exciting rotational population by a two-photon Raman process [20, 21]. Gas heating occurs from thermalization of the pumped rotational ensemble, which occurs over hundreds of picoseconds [22]. Previously it was shown [5] that after ~1 μs, a pressure-balanced quasi-equilibrium forms where the gas density profile is given by $\Delta N \approx -N_0 \Delta T/T_0$, where $N_0$ and $T_0$ are the background density and temperature and $\Delta T$ is the temperature increase from the laser absorption. Thus, the initial hole depth $|\Delta N|$ is proportional to the absorbed energy. The temperature profile, and therefore the density depression, then decays on millisecond timescales by thermal diffusion [5]. Even after many microseconds of diffusive spreading of the density hole [5], the peak depression is still proportional to the initial temperature, as verified in Fig. 1(a), in which a fluid code simulation [5] demonstrates the linear dependence of relative hole depth after 40 μs of evolution vs. the initial gas temperature change in nitrogen.

A diagram of the experimental setup is shown in Fig. 1. The laser pump consisted of a single pulse, double pulse, or a train of four 800nm, ~110 fs Ti:Sapphire pump pulses generated with a 4-pulse Michelson interferometer [23] ("pulse stacker"). The beam was focused at *f*/44 by a lens into a chamber filled with various gases. The vacuum FWHM of the beam waist was 33 μm. In all of the experiments, the laser was operated at 20 Hz to avoid cumulative effects of long time scale density depressions caused by previous pulses [5]. A continuous wave Helium-Neon laser at λ = 632.8 nm was used to interferometrically probe the 2D gas density. A plane at the pump beam waist was imaged into a folded wavefront interferometer and onto a CCD camera. Pump-induced changes in the gas density cause phase shifts $\Delta\varphi(x,y)$ in the $z$-propagating probe that are found by Fourier analysis of the interferogram [5]. Temporal gating of the probe pulse was achieved by triggering the CCD camera's minimum ~40 μs wide electronic shutter to include the pump pulse at the window's leading edge. Before phase extraction, 50 interferograms were averaged in order to improve the signal-to-noise ratio [24]. This reduced the RMS phase noise to ~6 mrad, enabling measurement of relative gas density changes $|\Delta N / N|$ as small as $10^{-4}$. The probe interaction length in the pump-heated gas is the pump beam confocal parameter, $L \approx 6.2$ mm. The gas density depression profile is given by $\Delta N(x,y) = (\lambda/2\pi)\Delta\varphi(x,y)N_{atm}/[(n_{atm}-1)L]$, where $N_{atm} \approx 2.47 \times 10^{19}$ cm$^{-3}$ is the molecular density at 1 atm and room temperature and $n_{atm}$ is the index of refraction of the test gas at 1 atm [25].

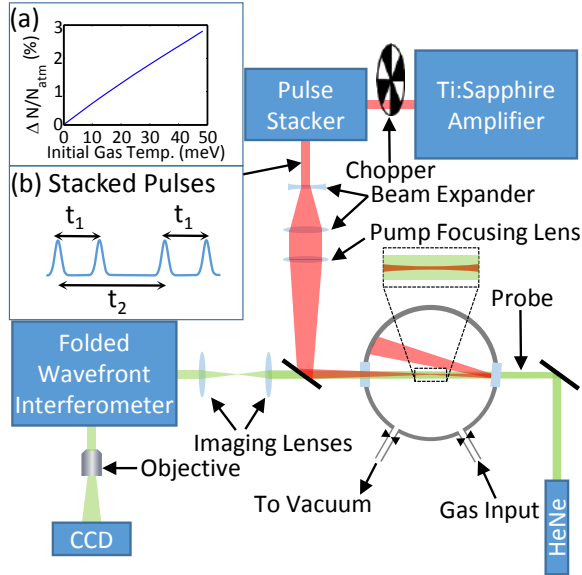

**Figure 1.** Experimental setup for measuring the 2D density profile of the rotationally excited gas at the pump beam focus. The chopper provides alternating pump on/off for background subtraction. (a) Simulation of hole depth vs. initial temperature showing that density hole depth is an excellent proportional measure of initial gas heating. (b) Scheme for ($t_1$, $t_2$) delay scan of pulses from pulse stacker.

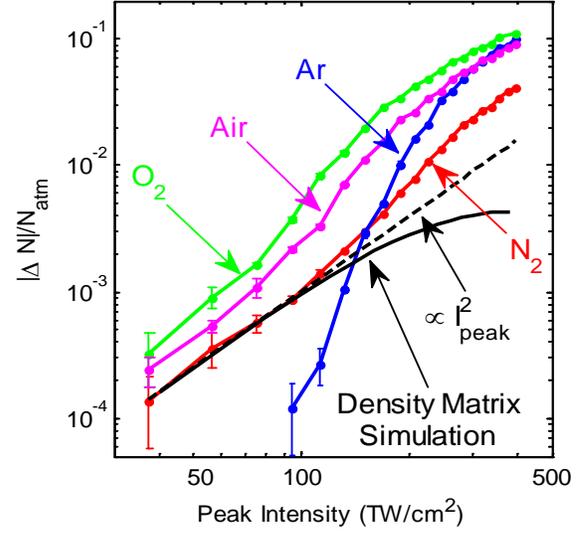

**Figure 2.** Relative density depression (proportional to heating) measured at 40 μs delay due to single-pulse (110 fs FWHM) rotational absorption versus pump intensity at focus. In argon, plasma generation from multiphoton ionization and tunneling is the only source of gas heating, whereas in diatomic molecules, rotational excitation enables nonlinear absorption below the ionization threshold. The solid black line shows a density matrix calculation of the rotational absorption in $N_2$ and the dashed line is a classical calculation using Eq. (3). The experimental points deviate from the density matrix simulation at higher intensities due to ionization and plasma absorption which is not modeled here.

In a preliminary experiment, we examined rotational absorption of single 110 fs pulses with energies ranging from 20 μJ to 500 μJ. Figure 2 shows the peak relative density hole depth $|\Delta N_{peak}|/N_{atm}$, measured at the center of the profile, as a function of the vacuum peak intensity. Measurements are shown for 1 atm $N_2$, $O_2$, Ar, and air. As discussed above, the relative hole depth $|\Delta N_{peak}|/N_{atm}$ is proportional to the laser energy absorption and initial gas temperature change. It is seen that its power dependence is quite different for the diatomic gases and Ar. For peak pump intensities of 40 TW/cm$^2$, below the ionization threshold of argon, we measured induced density depressions in all the diatomic gases, but none in argon to within our measurement uncertainty. At intensities below the ionization threshold, the energy absorbed by $N_2$ and $O_2$ has a roughly quadratic dependence on intensity, as expected for two-photon Raman absorption [8]. The curves deviate from the quadratic dependence at higher intensity, where absorption due to ionization strongly contributes. In all gases, saturation is observed at high pulse energy, which we attribute to the limiting of the laser intensity due to plasma defocusing [1]. Notably, our results show that at typical femtosecond filament clamping intensities of ~50 TW/cm$^2$, the greatly dominant source of laser energy deposition is rotational absorption and *not* ionization and plasma heating.

To calculate rotational absorption, we numerically solve for the evolution of the density matrix **ρ** describing the ensemble of molecules, which are assumed to be rigid rotors [11,26,27],

$$\frac{d\boldsymbol{\rho}}{dt} = -\frac{i}{\hbar}[H, \boldsymbol{\rho}], \qquad (1)$$

where [ ] denotes a commutator, and $H = H_0 + H_{opt}$ is the total Hamiltonian composed of $H_0 = \hat{L}^2/2I_M$ and the interaction between the optical field and the molecules, $H_{opt} = -\tfrac{1}{2}\mathbf{p}\cdot\mathbf{E}$. Here $\hat{L}$ is the rotational angular momentum operator, $I_M$ is the moment of inertia of the molecule, $\mathbf{p}$ is the induced dipole moment of the molecule, and $\mathbf{E}$ is the laser pulse electric field. For co-polarized optical pulses as used in our experiment, the interaction with the optical field only couples states with $\Delta j = \pm 2$ or 0 and $\Delta m = 0$. Initially at room temperature the rotational states are thermally populated, with the Boltzmann distribution peaking at approximately $j_{max} \sim 10$ in $N_2$. The initial density matrix is $\rho^{(0)}_{jmj'm'} = D_j \delta_{jj'} \delta_{mm'} \exp(-hcBj(j+1)/k_B T_0)/Z$, where $k_B$ is the Boltzmann constant, $B = \hbar/(4\pi c I_M)$ is the rotational constant (2.0 cm$^{-1}$ for $N_2$ [28]), $Z = \sum_k D_k(2k+1)\exp(-hcBk(k+1)/k_B T_0)$, and $D_j$ is a statistical weighting factor depending on the nuclear spin. For $N_2$, $D_j = 6$ for $j$ even and $D_j = 3$ for $j$ odd. The initial average rotational energy per molecule is $Tr(H_0\rho^{(0)}) = k_B T_0$, where $Tr$ is the trace operation. The change in average rotational energy $\Delta E$ per molecule (or the temperature change $k_B \Delta T$ of the molecular ensemble) induced by the pulse or pulse train is then given by

$$\Delta E = k_B \Delta T = Tr(H_0 \rho(t_f)) - Tr(H_0 \rho^{(0)}) =$$
$$= \sum_{j,m} hcBj(j+1)\rho_{(j,m),(j,m)}(t_f) - k_B T_0, \quad (2)$$

where $\rho(t)$ is evolved by Eq. (1) until time $t = t_f$ when the optical field from the pulse(s) is turned off. The pulses, individually or in a pulse train, are taken to be Gaussian in time.

The calculated rotational temperature change for a single pulse in $N_2$ is shown in Fig. 2 as a black solid line. The curve has been vertically shifted to match the experimentally measured hole depth in nitrogen. It matches well at low laser intensity $I$ where rotational absorption is expected to be proportional to $I^2$, but saturates at higher intensity as higher $j$-states become more separated in energy. The low intensity dependence can also be modeled classically as follows. For a classical rigid rotor, the torque on a molecule due to an optical field polarized at an angle $\theta$ with respect to the molecular axis is $\tfrac{1}{2}\Delta\alpha|\mathbf{E}|^2 \sin 2\theta$ where $\Delta\alpha$ is the molecular polarizability anisotropy. This can be used to show that the ensemble-averaged work done on a molecule in the limit of a single short pulse with fluence $F$ is

$$\Delta E_{classical} = \frac{16 F^2 \pi^2 (\Delta\alpha)^2}{15 c^2 I_M}. \quad (3)$$

This expression is plotted in Fig. 2 as a dashed line. The result agrees with the density matrix calculation at low intensities for a single pulse excitation of $N_2$ and $O_2$. We emphasize that the density matrix calculation predicts only the rotational absorption – at high intensities, the absorption is dominated by ionization and plasma heating as seen in the increasing deviation of the experimental points and simulation curves.

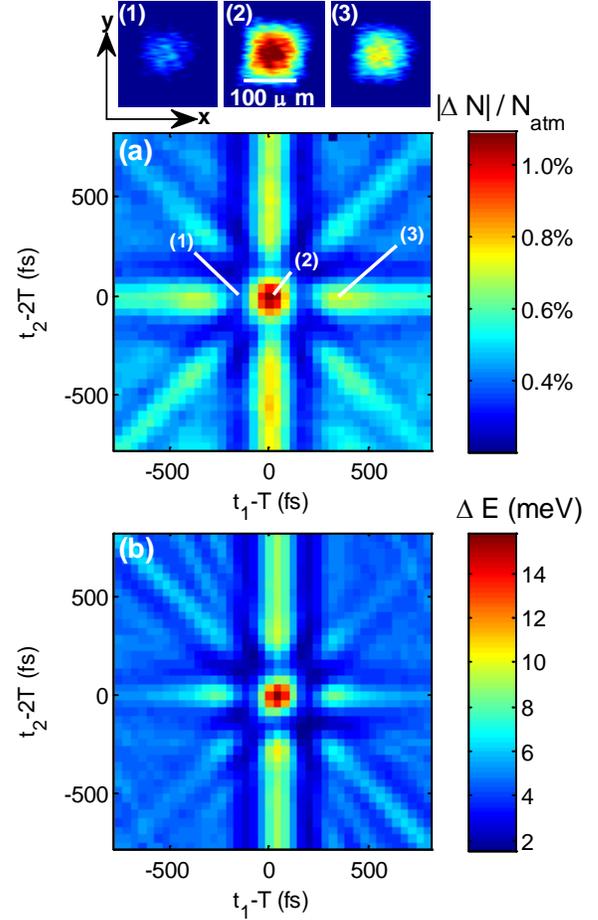

**Figure 3.** Rotational absorption as a function of time delays $t_1$ and $t_2$ in the pulse stacker. The images at the top show extracted density hole images for three delays, showing the varying depth of the hole. (a) Interferometric measurement of peak relative depth of gas density hole. The deepest gas density depression corresponds to the peak energy absorption predicted by the simulation. (b) Simulation of absorbed energy $\Delta E$, found by numerically solving Eqs. (1) and (2) and averaging along the pump beam's confocal parameter.

In the next experiment we investigated the effect of a 4-pulse train on the laser absorption and heating in nitrogen. With multiple pulses timed to match the rotational revival period, it is possible to strongly enhance the contribution of higher rotational states to the wavepacket ensemble [11]. Here we show directly that this translates into dramatically increased gas heating. The durations of pulses 1-4 were 110 fs, 110 fs, 120 fs and 110 fs, measured by a single shot autocorrelator, corresponding to vacuum peak intensities 61 TW/cm$^2$, 41 TW/cm$^2$, 41 TW/cm$^2$, 51 TW/cm$^2$. Note that if these pulses were coincident in time, the resulting single pulse would exceed the nitrogen ionization threshold. The pulse stacker time delays $t_1$ and $t_2$ were scanned by computer-controlled delay stages, so that the pulses arrived at $t = 0$, $t_1$, $t_2$, and $t_1+t_2$, as depicted in Fig. 1(b). We initially tuned the time separation between successive pulses to be $T$, the period of the first rotational revival. In nitrogen, $T = (2cB)^{-1} \approx 8.36$ ps, the time when the alignment revival crosses zero. Then, a fine 2D scan in $(t_1,t_2)$ of $|\Delta N_{peak}|/N_{atm}$ was performed with 40 fs steps.

Figure 3(a) shows the results of the pulse delay scan. Each point in the graph depicts the relative depth of the gas density hole at its center. The deepest hole, near $t_1 = T$ and $t_2 = 2T$, corresponds to the situation where each pulse in the train excites the molecules at the full revival from the previous pulse. Other features in the plot can be understood as resonances involving fewer pulses. The vertical bar ($t_1 = T$, $t_2 = 2T + \Delta t$) is where the first two pulses and second two pulses are resonant, but the delay between the second and third pulses is not. The horizontal bar ($t_1 = T +\Delta t$, $t_2 = 2T$) is where the first and third pulses are resonant (at the second full revival of the first pulse), and the same with the second and fourth pulses. The diagonals correspond to resonances between two pulses in the train. The southwest to northeast diagonal ($t_1 = T + \Delta t$, $t_2 = 2T + \Delta t$) is where the second and third pulses are resonant. The southeast to northwest diagonal ($t_1 = T + \Delta t$, $t_2 = 2T - \Delta t$) is where the first and fourth pulses are resonant (at the second half revival). The maximum depth of the gas density hole induced by the 4-pulse train is ~6 times greater than the minimum in Fig. 3(a), which is similar to that induced by a single pulse.

Figure 3(b) shows a simulation, using Eqs. (1) and (2), of the absorbed energy in $N_2$ as a function of the $(t_1, t_2)$ scan of a 4-pulse train using the experimental pulse parameters as inputs. The simulations predict dramatic ionization-free heating of as much as 30 meV/molecule ($\Delta T \sim 350$ K) at the pump beam waist, which is only matched by a ~100 eV filament plasma (with electron density $2\times10^{16}$ cm$^{-3}$ [29]). The plot shows an axial average peak heating of 15 meV/molecule along the 6.2 mm pump confocal parameter. Comparison to Fig. 3(a) shows very good agreement between experiment and theory, with the resonance bars in the simulation decaying somewhat faster from the heating peak than in the experiment, an effect we are investigating. A similar experiment and simulation were performed for $O_2$ gas, likewise with good agreement.

So far we showed that it is possible to coherently excite a rotational wavepacket ensemble with a sequence of pulses separated by a full revival period, leading to strong heating of a gas of diatomic molecules. However, it is also possible to first excite the ensemble and then de-excite it well within the decoherence time over which it would normally thermalize and fully heat the gas. To show this we used two pulses out of the pulse stacker. The first pulse was used to excite the ensemble. We scanned the arrival time of the second pulse, $t_1$, near $T/2 \sim 4.2$ ps, the half-revival period of nitrogen. Figure 4 shows the measured depth of the density hole reduced by ~65% at the half-revival delay, while the simulation shows an absorption reduction of ~82%. In essence, energy from the first pulse invested in the wavepacket ensemble is coherently restored to the second pulse. Viewed alternatively, the $T/2$-delayed second pulse acts as an out-of-phase kick to suppress the molecular alignment induced by the first pulse, in contrast to the $T$-delayed pulses which act as in-phase kicks to enhance alignment.

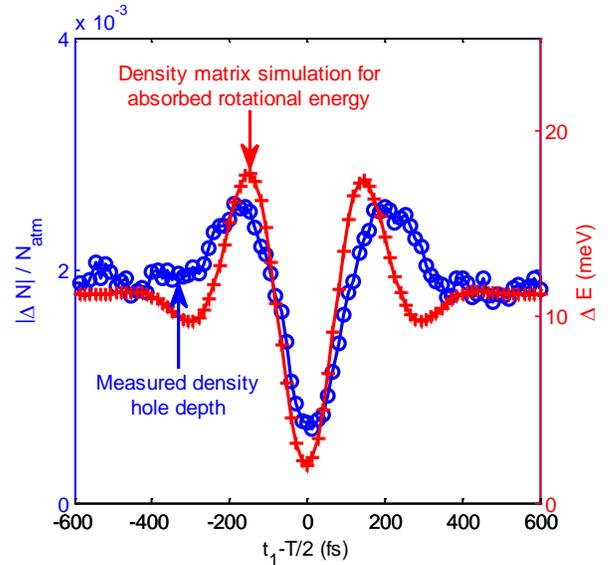

**Figure 4.** Reduction of rotational heating using two pulses. The measured change in gas density is shown as blue circles as a function of the time delay $t_1$ between two pulses spaced near the half revival $T/2$=4.16 ps in $N_2$. The hole depth is reduced by ~65%. The red +'s show the rotational energy change per molecule from solving Eqs. (1) and (2) for varying $t_1$.

In summary, we have measured the dramatic gas hydrodynamic response to coherent excitation and de-excitation of a molecular rotational wavepacket ensemble, at peak laser intensities well below the ionization threshold. The laser absorption and gas heating is significantly enhanced by using a 4-pulse train with pulses separated by the molecular rotational revival time. Heating is strongly suppressed by coherently de-exciting the molecular ensemble using pulses spaced by a half-revival. The femtosecond sensitivity to pulse train timing of gas heating and heating suppression is well predicted by density matrix simulations of the evolution of the wavepacket ensemble. Our results make possible the fine quantum control of gas density profiles using non-ionizing laser pulses. Such profile modification, at both near and remote locations, has a range of exciting applications including the refractive index control of high power optical pulse propagation in air [5-7].

The authors acknowledge useful discussions with J. Palastro. This work is supported by the Air Force Office of Scientific Research, the Office of Naval Research, the Department of Energy, and the National Science Foundation.

**References**

[1] A. Couairon and A. Mysyrowicz, Phys. Rep. **441**, 47 (2007).
[2] J. Yu, D. Mondelain, J. Kasparian, E. Salmon, S. Geffroy, C. Favre, V. Boutou, and J.-P. Wolf, Appl. Opt. **42**, 7117 (2003).
[3] F. Vidal, D. Comtois, C.-Y. Chien, A. Desparois, B. La Fontaine, T. W. Johnston, J.-C. Kieffer, H. P. Mercure, H. Pepin, and F. A. Rizk, IEEE Trans. Plasma Sci. **28**, 418 (2000).
[4] S. Tzortzakis, B. Prade, M. Franco, A. Mysyrowicz, S. Huller, and P. Mora, Phys. Rev. E **64**, 057401 (2001).
[5] Y.-H. Cheng, J. K. Wahlstrand, N. Jhajj, and H. M. Milchberg, Opt. Express **21**, 4740 (2013).
[6] N. Jhajj, Y.-H. Cheng, J. K. Wahlstrand, and H. M. Milchberg, Opt. Express **21**, 28980 (2013).
[7] N. Jhajj, E. W. Rosenthal, R. Birnbaum, J. K. Wahlstrand, and H. M. Milchberg, arXiv:1311.1846.
[8] D. V. Kartashov, A. V. Kirsanov, A. M. Kiselev, A. N. Stepanov, N. N. Bochkarev, Y. N. Ponomarev, and B. A. Tikhomirov, Opt. Express **14**, 7552-7558 (2006).
[9] A. M. Kiselev, Yu. N. Ponomarev, A. N. Stepanov, A. B. Tikhomirov, and B. A. Tikhomirov, Quantum Electron. **41**, 976-979 (2011).
[10] Tz. B. Petrova, H. D. Ladouceur, and A. P. Baronavski, Phys. Plasmas **15**, 053501 (2008).
[11] A. York and H. M. Milchberg, Opt. Express **16**, 10557 (2008)
[12] J. P. Cryan, P. H. Bucksbaum, and R. N. Coffee, Phys. Rev. A **80**, 063412 (2009).
[13] S. Varma, Y.-H. Chen, J. P. Palastro, A. B. Fallahkhair, E. W. Rosenthal, T. Antonsen, and H. M. Milchberg, Phys. Rev. A **86**, 023850 (2012).
[14] J. P. Palastro, T. M. Antonsen, Jr., and H. M. Milchberg, Phys. Rev. A **86**, 033834 (2012).
[15] S. Zhdanovich, C. Bloomquist, J. Floss, I. Sh. Averbukh, J. W. Hepburn, and V. Milner, Phys. Rev. Lett. **109**, 043003 (2012).
[16] D. M. Villeneuve, S. A. Aseyev, P. Dietrich, M. Spanner, M. Yu. Ivanov, and P. B. Corkum, Phys. Rev. Lett. **85**, 542-545 (2000).
[17] L. Yuan, C. Toro, M. Bell, and A. S. Mullin, Faraday Discuss. **150**, 101 (2011).
[18] S. Zhdanovich, A. A. Milner, C. Bloomquist, J. Floss, I. Sh. Averbukh, J. W. Hepburn, and V. Milner, Phys. Rev. Lett. **107**, 243004 (2011).
[19] U. Steinitz, Y. Prior, and I. S. Averbukh, Phys. Rev. Lett. **109**, 033001 (2012).
[20] H. Stapelfeldt and T. Seideman, Rev. Mod. Phys. **75**, 543557 (2003).
[21] C. H. Lin, J. P. Heritage, T. K. Gustafson, R. Y. Chiao, and J. P. McTague, Phys. Rev. A **13**, 813 (1976).
[22] D. R. Miller and R. P. Andres, J. Chem. Phys. **46**, 3418 (1967).
[23] C. W. Siders, J. L. W. Siders, A. J. Taylor, S. G. Park, and A. M. Weiner, Appl. Opt. **37**, 5302 (1998).
[24] Y.-H. Chen, S. Varma, I. Alexeev, and H. M. Milchberg, Opt. Express **15**, 7458 (2007).
[25] K. P. Birch, J. Opt. Soc. Am. A **8**, 647 (1991).
[26] S. Ramakrishna and T. Seideman, Phys. Rev. Lett. **95**, 113001 (2005).
[27] Y.-H. Chen, S. Varma, A. York, and H. M. Milchberg, Opt. Express **15**, 11341 (2007).
[28] K. P. Huber and G. Herzberg, *Molecular Spectra and Molecular Structure: IV. Constants of Diatomic Molecules* (van Nostrand Reinhold, New York, 1979).
[29] Y.-H. Chen, S. Varma, T. M. Antonsen, and H. M. Milchberg, Phys. Rev. Lett. **105**, 215005 (2010).